\documentclass[twocolumn]{aastex63}      
\usepackage{booktabs}    
\usepackage{bibunits}
\usepackage{enumerate}
\usepackage{amsmath, amssymb}
\usepackage{rotating}
\usepackage{float}
\usepackage{threeparttable} 

\shortauthors{Hou et al.}

\begin{document}
\title{Characterizing the Gamma-ray Emission Properties of the Globular Cluster M5 with the Fermi-LAT}

\correspondingauthor{X. Hou, D. F. Torres}
\email{xhou@ynao.ac.cn, dtorres@ice.csic.es}

\author[0000-0003-0933-6101]{X. Hou}
\affiliation{Yunnan Observatories, Chinese Academy of Sciences, Kunming 650216, People's Republic of China}
\affiliation{Key Laboratory for the Structure and Evolution of Celestial Objects, Chinese Academy of Sciences, Kunming 650216, People's Republic of China}
\affiliation{Center for Astronomical Mega-Science, Chinese Academy of Sciences, Beijing 100012, People's Republic of China}

\author[0000-0003-2839-1325]{W. Zhang}
\affiliation{Institute of Space Sciences (ICE, CSIC), Campus UAB, 08193 Barcelona, Spain}
\affiliation{Institut d’Estudis Espacials de Catalunya (IEEC), 08034 Barcelona, Spain}

\author[0000-0003-1307-9435]{P. C. C. Freire}
\affiliation{Max-Planck Institut f{\"u}r Radioastronomie, Auf dem H{\"u}gel 69, D-53121 Bonn, Germany}

\author[0000-0002-1522-9065]{D. F. Torres}
\affiliation{Institute of Space Sciences (ICE, CSIC), Campus UAB, 08193 Barcelona, Spain}
\affiliation{Institut d’Estudis Espacials de Catalunya (IEEC), 08034 Barcelona, Spain}
\affiliation{Institució Catalana de Recerca i Estudis Avançats (ICREA), E-08010 Barcelona, Spain}

\author[0000-0002-8784-2977]{J. Ballet}
\affiliation{Universit\'e Paris Saclay and Universit\'e Paris Cit\'e, CEA, CNRS, AIM, F-91191 Gif-sur-Yvette, France}

\author[0000-0002-7833-0275]{D. A. Smith}
\affiliation{Laboratoire d’Astrophysique de Bordeaux, Universit\'e de Bordeaux, CNRS, B18N, all\'ee Geoffroy Saint-Hilaire, F-33615 Pessac, France}

\author[0000-0002-2771-472X]{T. J. Johnson}
\affiliation{College of Science, George Mason University, Fairfax, VA 22030, resident at Naval Research Laboratory, Washington, DC 20375, USA}

\author[0000-0002-0893-4073]{M. Kerr}
\affiliation{Space Science Division, Naval Research Laboratory, Washington, DC 20375-5352, USA}

\author[0000-0002-4377-0174]{C. C. Cheung}
\affiliation{Space Science Division, Naval Research Laboratory, Washington, DC 20375-5352, USA}

\author[0000-0002-9049-8716]{L. Guillemot}
\affiliation{Laboratoire de Physique et Chimie de l’Environnement et de l’Espace – Universit\'e d’Orl\'eans/CNRS, F-45071 Orl\'eans Cedex 02, France}
\affiliation{Station de radioastronomie de Nan\c{c}ay, Observatoire de Paris, CNRS/INSU, F-18330 Nan\c{c}ay, France}

\author[0000-0003-1720-9727]{J. Li}
\affiliation{CAS Key Laboratory for Research in Galaxies and Cosmology, Department of Astronomy, University of Science and Technology of China, Hefei 230026, People's Republic of China}
\affiliation{School of Astronomy and Space Science, University of Science and Technology of China, Hefei 230026, People's Republic of China}

\author[0000-0001-8539-4237]{L. Zhang}
\affiliation{National Astronomical Observatories, Chinese Academy of Sciences, Beijing 100101, People's Republic of China}
\affiliation{Centre for Astrophysics and Supercomputing, Swinburne University of Technology, P.O. Box 218, Hawthorn, VIC 3122, Australia}

\author[0000-0001-6762-2638]{A. Ridolfi}
\affiliation{INAF – Osservatorio Astronomico di Cagliari, Via della Scienza 5, I-09047 Selargius (CA), Italy}
\affiliation{Max-Planck Institut f{\"u}r Radioastronomie, Auf dem H{\"u}gel 69, D-53121 Bonn, Germany}


\author[0000-0002-3386-7159]{P. Wang}
\affiliation{National Astronomical Observatories, Chinese Academy of Sciences, Beijing 100101, People's Republic of China}
\affiliation{Institute for Frontiers in Astronomy and Astrophysics, Beijing Normal University, Beijing 102206, People's Republic of China}

\author[0000-0003-3010-7661]{D. Li}
\affiliation{National Astronomical Observatories, Chinese Academy of Sciences, Beijing 100101, People's Republic of China}
\affiliation{University of Chinese Academy of Sciences, Chinese Academy of Sciences, Beijing 100101, People's Republic of China}
\affiliation{Zhijiang Lab, Hangzhou 311121, People’s Republic of China}

\author[0000-0002-5381-6498]{J. Yuan}
\affiliation{Xinjiang Astronomical Observatory, Chinese Academy of Sciences, Urumqi, Xinjiang 830011, People’s Republic of China}

\author[0000-0002-9786-8548]{N. Wang}
\affiliation{Xinjiang Astronomical Observatory, Chinese Academy of Sciences, Urumqi, Xinjiang 830011, People’s Republic of China}


\begin{abstract}
We analyzed the globular cluster M5 (NGC 5904) using 15 yr of gamma-ray data from the Fermi Large Area Telescope (LAT). Using rotation ephemerides generated from Arecibo and FAST radio telescope observations, we searched for gamma-ray pulsations from the seven millisecond pulsars (MSPs) identified in M5. We detected no significant pulsations from any of the individual pulsars.
In addition, we searched for possible variations of the gamma-ray emission as a function of orbital phase for all six MSPs in binary systems, but we did not detect any significant modulations. 
The gamma-ray emission from the direction of M5 is well described by an exponentially cutoff power-law spectral model, although other models cannot be excluded. The phase-averaged emission is consistent with being steady on a time scale of a few months. We estimate the number of MSPs in M5 to be between 1 and 10, using the gamma-ray conversion efficiencies for well-characterized gamma-ray MSPs in the Third Fermi LAT of Gamma-ray Pulsars, suggesting that the sample of known MSPs in M5 is (nearly) complete, even if it is not currently possible to rule out a diffuse component of the observed gamma rays from the cluster.
\end{abstract}


\section{Introduction}
\label{sec:introduction}

Globular clusters (GCs) are the oldest and densest stellar systems bound by gravity. Due to the high stellar density \citep[$>1000$ pc$^{-3}$;][]{Sollima2017} and frequent dynamical interactions between stars in GCs, the formation rate per unit mass of low-mass X-ray binaries (LMXBs) is orders of magnitude higher in GCs than in the Galactic field \citep{Clark1975,Katz1975}. LMXBs are more abundant in GCs as a natural consequence of such a dynamical formation scenario, and a linear correlation between the number of LMXBs in GCs and the stellar encounter rate $\Gamma_{c}$ has been expected and confirmed by observations of GCs \citep{Gendre2003,Pooley2003,Menezes2023}.
%

Millisecond pulsars (MSPs; usually defined as those having spin period $P\le30$ ms) are generally believed to be descendants of LMXBs \citep[e.g.,][]{Alpar1982,Bhattacharya1991}. Of the 305 pulsars detected in radio in 40 GCs in the Milky Way (MW) halo,\footnote{\url{https://www3.mpifr-bonn.mpg.de/staff/pfreire/GCpsr.html}} 80\% are MSPs. In comparison, 427 known MSPs are not associated with GCs,\footnote{\url{http://astro.phys.wvu.edu/GalacticMSPs/GalacticMSPs.txt}} only 10\% of the known Galactic pulsar population.
A positive correlation between the MSP population in GCs and $\Gamma_{c}$ has also been reported \citep[e.g.,][]{Hui2010,Bahramian2013}, which provided evidence for the dynamical origin of MSPs as had long been predicted, given the close relation between MSPs and LMXBs.


GCs have been established as a class of gamma-ray emitters using data from the Large Area Telescope \citep[LAT;][]{Atwood2009} on board the \textit{Fermi Gamma-ray Space Telescope\footnote{\url{http://fermi.gsfc.nasa.gov/ssc}}} launched in 2008. Up to now, gamma rays coincident with the directions of about 39 GCs have been reported \citep{Abdo2010,Kong2010,Tam2011,Zhou2015,Zhang2016,Zhangp2022,Lloyd2018,Menezes2019,4FGL,Yuan2022a,Yuan2022b,Menezes2023}, all of which are listed in the \cite{Harris1996} catalog\footnote{2010 edition: \url{https://physics.mcmaster.ca/\~harris/mwgc.dat}} of MW GCs.
As a main class of LAT gamma-ray sources, MSPs are reasonably thought 
to be responsible for the collective gamma-ray emission from GCs, as first predicted by \cite{Chen1991} and suggested by the spectral similarities between the observed gamma-ray MSPs and GCs. 
Indeed, evidence of correlation between the gamma-ray luminosity $L_{\gamma}$ and $\Gamma_{c}$ established by various studies has provided support to the dynamical formation of MSPs and the MSP origin of gamma rays in GCs \citep{Abdo2010,Hui2011,Bahramian2013,Hooper2016,Zhang2016,Menezes2019,Menezes2023,Feng2024}.
Recent analyses show that all GCs are point-like sources in gamma rays, implying that MSPs are mostly concentrated in their cores
\citep{Menezes2019,Menezes2023}.

Exceptionally, gamma-ray pulsations from individual MSPs have also been reported in three GCs: PSR J1823$-$3021A in NGC 6624 \citep{Freire2011}, PSR B1821$-$24 in NGC 6626 \citep[M28,][]{Johnson2013,Wu2013}, and PSR J1835$-$3259B in NGC 6652 \citep{Zhangp2022}. 
The first two are isolated MSPs, while the third one is in a binary system \citep{2022A&A...664A..54G}. They are all energetic pulsars with relatively high spin-down power and are bright in gamma rays, indicating that they probably dominate the gamma-ray emission from the host GCs.
We have been unable to confirm the gamma-ray pulsations from PSR J1717+4308A in NGC 6341 (M92) reported by \cite{pZhang2023}.

\begin{table*}[htbp]
\scriptsize
\begin{center}
\caption{Basic properties of the seven MSPs in M5}
\label{tab:M5info}
\begin{tabular}{lccccccccc}
\toprule %
\midrule %
Name  &R.A.  &Decl.   &$P$  & $P_{\rm orb}$ & Eccent.$^{a}$ & $\dot E$$^{b}$   & X-Ray  &Optical &Comment \\
&(deg)   &(deg)   &(ms)  & (day)   &  &(10$^{34}$~erg~s$^{-1}$)  &Counterpart   &Counterpart &   \\
\midrule %
M5A  & 229.6388   & 2.0910   & 5.55	    & ...       & ...  &$<1.56$   & No & No & Isolated \\
M5B  & 229.6311   & 2.0876	 & 7.95	    & 6.858     & 0.138  &$<0.25$  & No & No  &Heavy, not likely edge-on \\
M5C  & 229.6366	  & 2.0799	  & 2.48	& 0.687    & 0    &$<12.36$    &Yes  & Yes  & Eclipsing BW\\
M5D  & 229.6268   & 2.0833	 & 2.99	    & 1.222	    & lower than M5B     &0.28-4.98  &Yes & Yes  &He WD companion\\
M5E  & 229.6388   & 2.0772	 & 3.18	   & 1.097	    & lower than M5B     &$<5.27$  &Yes & Yes &He WD companion\\
M5F  & 229.6350	   & 2.0867	 & 2.65	   & 1.610      & lower than M5B    &$<8.30$  & No & Yes   &He WD companion\\
M5G  & 229.6197   & 2.0875	 & 2.75	    & 0.114    & 0   &0.04-3.34    & Yes  & No  & Noneclipsing BW\\
\bottomrule %
\multicolumn{10}{p{17cm}}{$^{a}$ The eccentricities of M5C and M5G have been set to 0 based on the assumption that the orbits of BW pulsars are circular owing to tidal dissipation \citep[see Table 1 in ][]{lZhang2023}.} \\
\multicolumn{10}{p{17cm}}{$^{b}$ Spin-down powers $\dot E$ have been calculated based on the intrinsic spin-down rate $\dot P_{\rm int}$ upper limit corrected for accelerations caused by the gravitational field of the GC for the line of sight of each pulsar \citep[see Table 2 in ][]{lZhang2023}.}
\end{tabular}
\end{center}
\end{table*}

M5 (NGC 5904) is a bright GC (visual magnitude $V\approx5.6$) with a small $\Gamma_{c}$, which is also small when we divide it by the number of stars in the cluster, i.e., the encounter rate per formed binary, $\gamma_{b}$ \citep{Verbunt_Freire2014}. 
The cluster is at a distance of 7.48$\pm$0.60 kpc {with a half-mass radius of 5.6 pc.\footnote{\url{https://people.smp.uq.edu.au/HolgerBaumgardt/globular/}}
%
Radio observations using Arecibo and FAST have resulted in the detection of seven MSPs in M5 \citep{Anderson1997,Mott2003,Hessels2007,Pan2021,lZhang2023}. The first discoveries were then known as B1516+02A and J1518+0204B; they are now known as PSRs J1518+0204A and J1518+0204B. The following discoveries are known as J1518+0204C - J1518+0204G, and we will refer to these pulsars as M5A, M5B, M5C, M5D, M5E, M5F, and M5G, respectively. 
Improved timing solutions for all these pulsars have been derived based on observations using the two telescopes \citep{lZhang2023}.

The basic properties of the seven MSPs in M5 are presented in Table~\ref{tab:M5info}.
M5A is an isolated MSP, while the other six are in binary systems. Based on optical observations, the companions of M5D, M5E, and M5F are very likely low-mass He white dwarfs (WDs). 
M5C, M5D, M5E, and M5G have X-ray counterparts detected by \textit{Chandra},
from which thermal X-ray emission is observed. 
M5C (eclipsing) and M5G (noneclipsing) are black widow (BW) systems that show no or little nonthermal X-ray emissions, indicating that the intrabinary shock produces weak synchrotron radiation. 
The improved measurement of the periastron advance rate $\dot \omega=0\fdg01361$ yr$^{-1}$ of M5B is compatible
with a heavy neutron star (NS), consistent with previous studies. 
Although its inclination is not well constrained, M5B is probably not edge-on \citep{lZhang2023}.

M5 was initially proposed as a gamma-ray emitter by \cite{Zhou2015} with a marginal detection ($3.2\sigma$) using 6 yr of LAT Pass 7 Reprocessed data and later confirmed by \cite{Zhang2016} with $4.4\sigma$ using 7 yr of Pass 8 data \citep{Atwood2013,Bruel2018}. The Fermi-LAT Fourth Source Catalog \citep[4FGL;][]{4FGL} first associated M5 with the LAT source 4FGL J1518.8+0203 (6.7$\sigma$) using 8 yr of Pass 8 data.

In this paper, we performed Fermi-LAT analysis of M5 to investigate 
its gamma-ray emission properties using an updated data set and the most recent LAT catalog. 
From Table~\ref{tab:M5info}, the angular separations between the seven MSPs are $\sim 0\fdg01-0\fdg02$; thus, the LAT angular resolution ($\sim 0\fdg1$, Figure 1 in 4FGL) is insufficient to spatially separate the individual MSPs.
Motivated by this, we have performed timing analysis in addition to spectral analysis taking advantage of the most updated timing solutions for the seven MSPs in M5.
Either detection or nondetection of individual pulsars usefully constrains the number of MSPs in M5 and the GC gamma-ray emission models and serves as a useful test of the dynamical formation scenario of GC MSPs. 
We describe the data set and analysis methods in Section \ref{dataset}, and then we present the analysis results in Section \ref{result}. Finally, we discuss the implications of the results and conclude in Section \ref{discuss}.

\section{Data set and analysis methods}
\label{dataset}

\begin{figure*}[htbp]
    \centering 
        \includegraphics[width=0.49\linewidth]{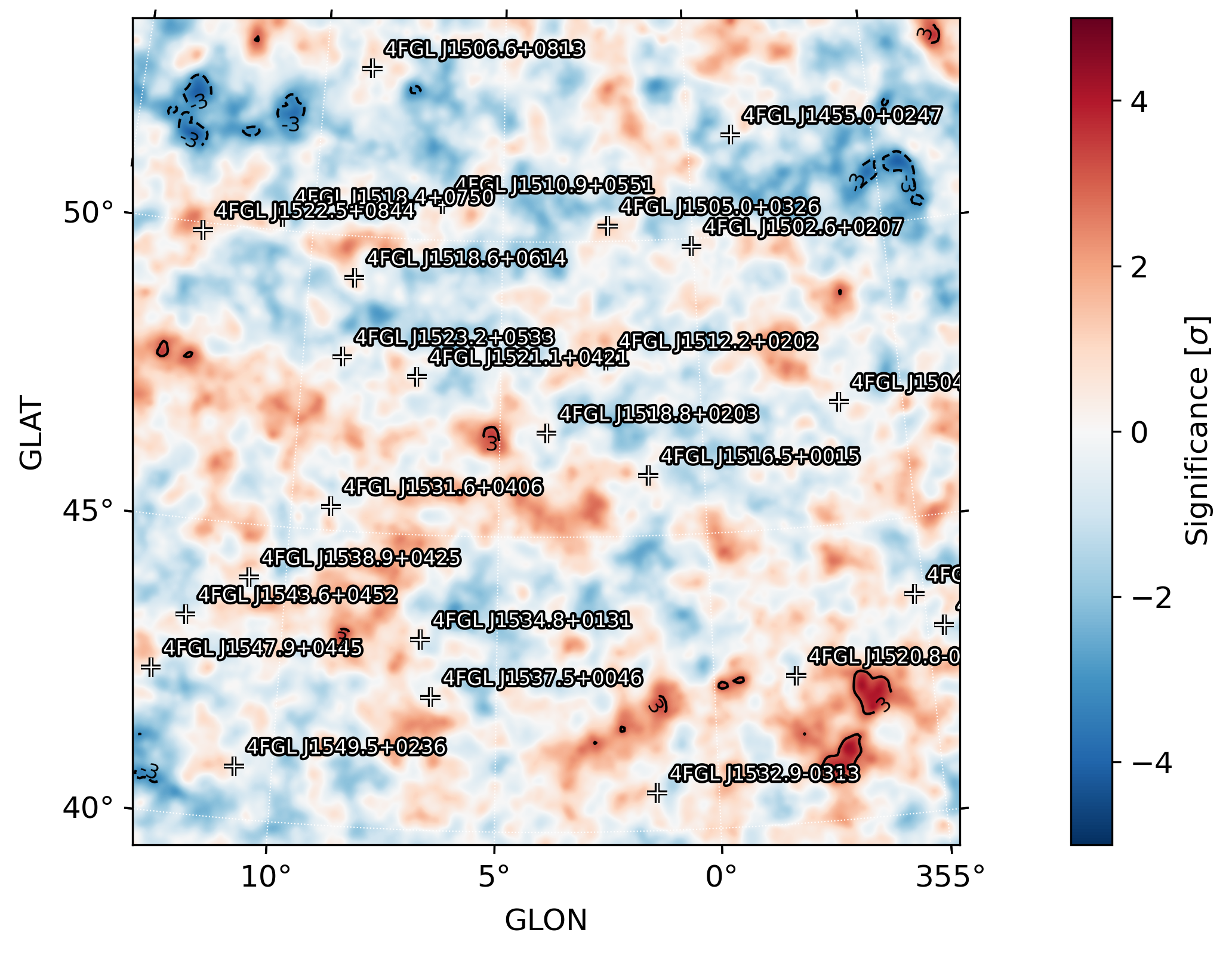}
        \includegraphics[width=0.49\linewidth]{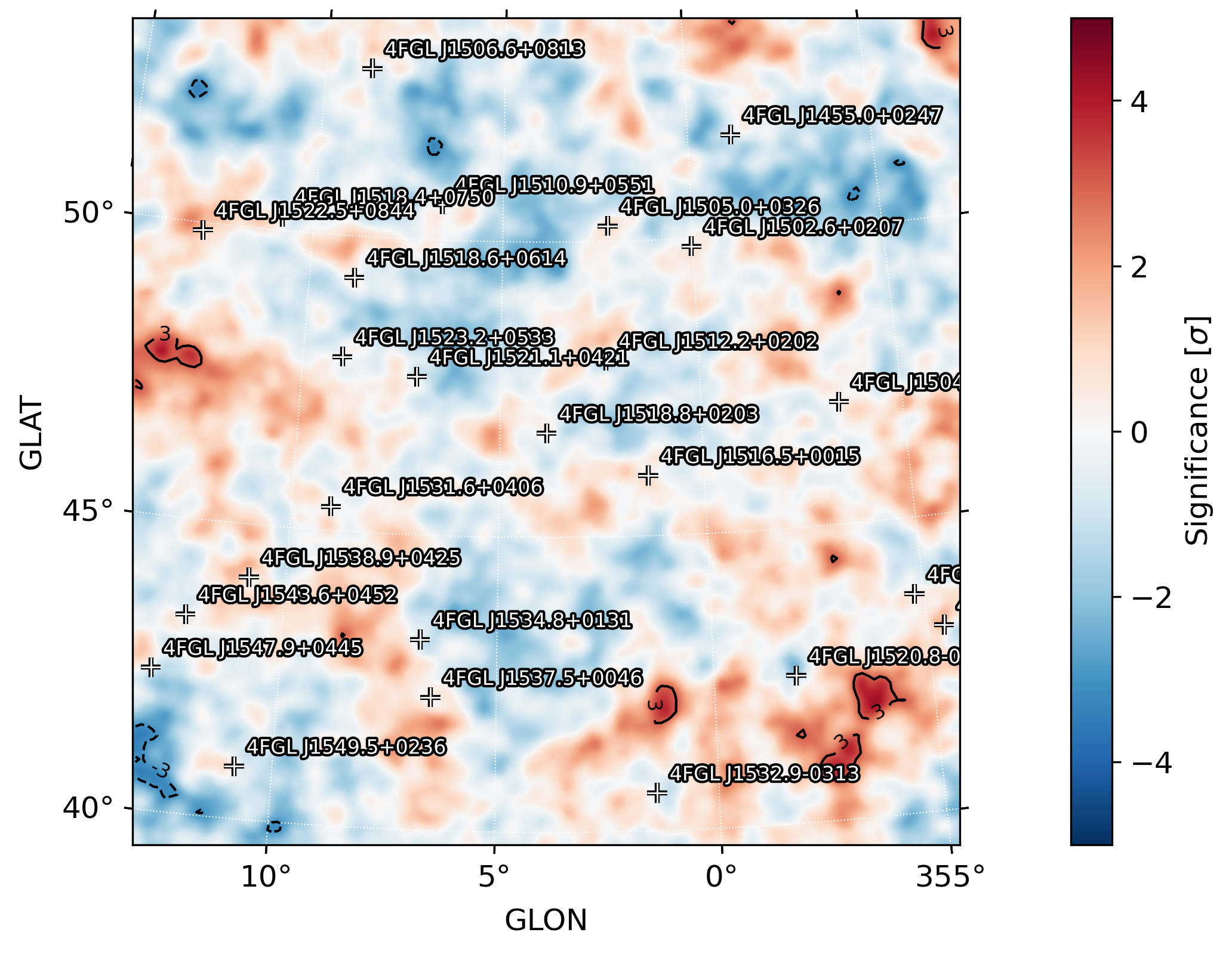}
        \caption{Residual maps of M5 for the full ROI in $\sigma$ units (in Galactic coordinates). Black plus signs are 4FGL-DR4 sources included in the ROI. Left: In the 1$-$500 GeV band for the localization of M5. Right: In the 0.1$-$500 GeV band for the spectral fit for M5.}
    \label{fig:residualmap}
\end{figure*}

According to the latest 4FGL-DR4 source list \citep[gll\_psc\_v32.fit;\footnote{\url{https://fermi.gsfc.nasa.gov/ssc/data/access/lat/14yr_catalog/}}][]{4FGL-DR4}, M5 is associated with 4FGL J1518.8+0203, which is 4.1$'$ away from the M5 center and has a gamma-ray significance of 6.8$\sigma$.
We used 15 yr (2008 August 4$-$2023 August 4) of Pass 8 data from the Fermi-LAT within $10^\circ$ around the M5 center $(\alpha,\delta)=(229\fdg6384, 2\fdg0810$) in the J2000 frame.\footnote{\url{http://simbad.u-strasbg.fr/simbad/}} 
SOURCE class events in the energy range of 0.1$-$500 GeV have been selected and the standard event filter ``\texttt{DATA\_QUAL>0 \&\& LAT\_CONFIG==1}'' has been applied to get data of good quality.\footnote{\url{https://fermi.gsfc.nasa.gov/ssc/data/analysis/scitools/data\_preparation.html}} 
To avoid contamination from solar flares and gamma-ray bursts (GRBs), we have excluded time intervals when solar flares and GRBs occurred \citep{4FGL-DR3,4FGL-DR4}. 

To reduce the contamination from the low-energy Earth limb emission, we followed 
similar point-spread function (PSF) and zenith-angle cuts adopted in the LAT 4FGL catalog \citep{4FGL}. 
Specifically, in the 0.1$-$0.3 GeV band, only PSF2 and PSF3 events were selected with zenith angles $<90^\circ$; in the 0.3$-$1 GeV band, PSF1, PSF2, and PSF3 events were selected with zenith angles $<100^\circ$; and in the $1-500$ GeV band, all events (PSF0, PSF1, PSF2, and PSF3) with zenith angles $<105^\circ$ were used. 
This reduces the contribution of the Earth limb contamination to the total background to less than 10\%. 

We built a spatial-spectral model for M5 by including the 4FGL-DR4 sources within $20^\circ$ around the M5 center.
The Galactic interstellar emission model (``gll\_iem\_v07.fits'') and the isotropic emission spectrum (``iso\_P8R3\_SOURCE\_V3\_v1.txt''), which takes into account the extragalactic emission and the residual instrumental background,\footnote{\url{https://fermi.gsfc.nasa.gov/ssc/data/access/lat/BackgroundModels.html}} were also included.
Energy dispersion has been taken into account by adding two extra energy bins except for the isotropic component. 
We have performed a summed likelihood analysis in a $14^\circ \times 14^\circ$ region of interest (ROI) around M5.
The significance of a given source in the model is characterized by the test statistic (TS).\footnote{TS $=2(\log \mathcal{L}-\log \mathcal{L}_{0})$, where $\log \mathcal{L}$ and $\log \mathcal{L}_{0}$ are the logarithms of the maximum likelihood of the complete source model and of the model without the target source included, respectively \citep{mattox96}.} 
In this work, we used the $\mathtt{fermipy}$\footnote{\url{http://fermipy.readthedocs.io/en/latest/index.html}} package \citep[v1.2.0;][]{Wood2017} in which Fermitools\footnote{\url{https://fermi.gsfc.nasa.gov/ssc/data/analysis/software}} (v2.2.0) is integrated.

\section{Analysis results}
\label{result}
\begin{figure*}[htbp]
    \centering 
        \includegraphics[width=0.49\linewidth]{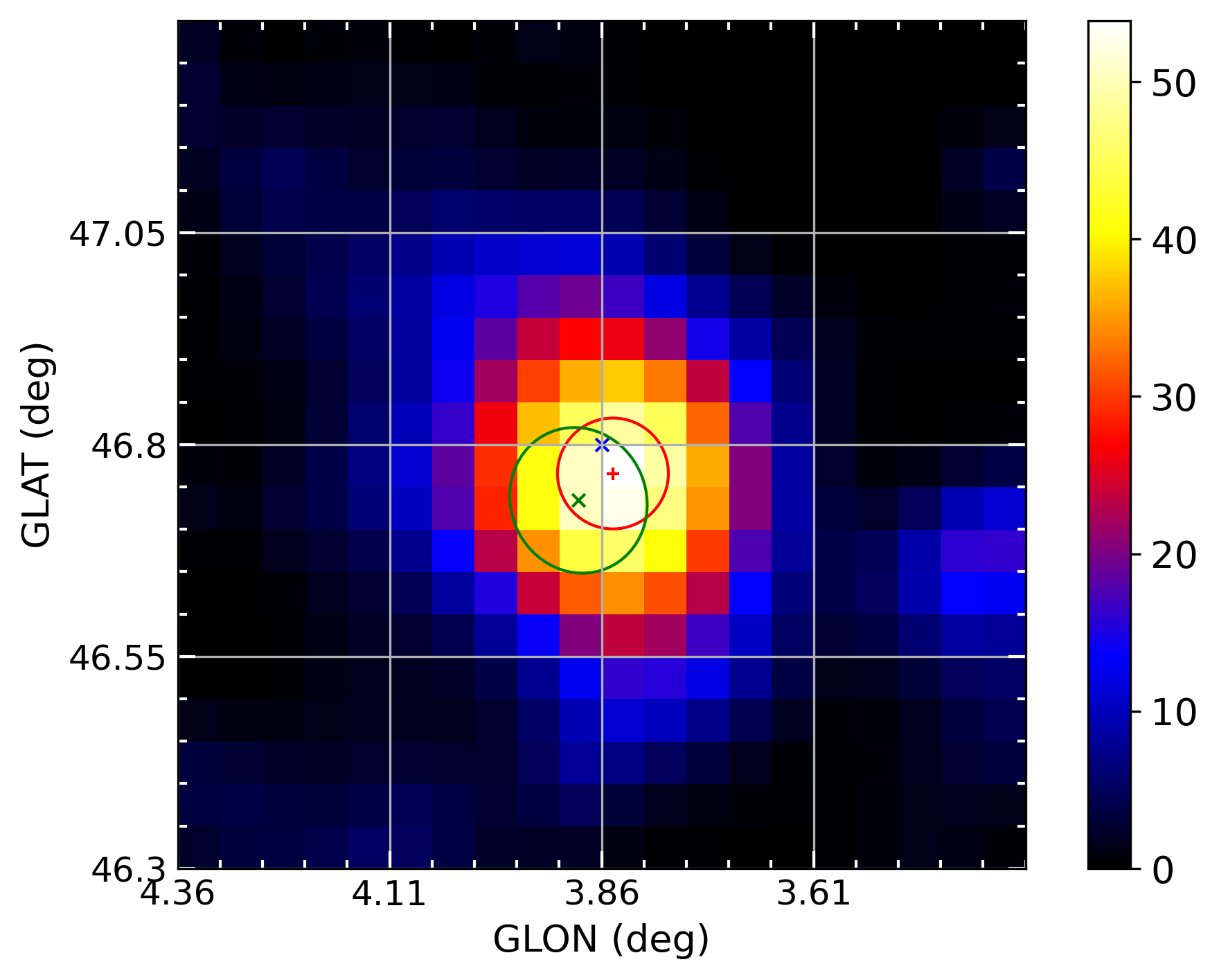}
        \includegraphics[width=0.49\linewidth]{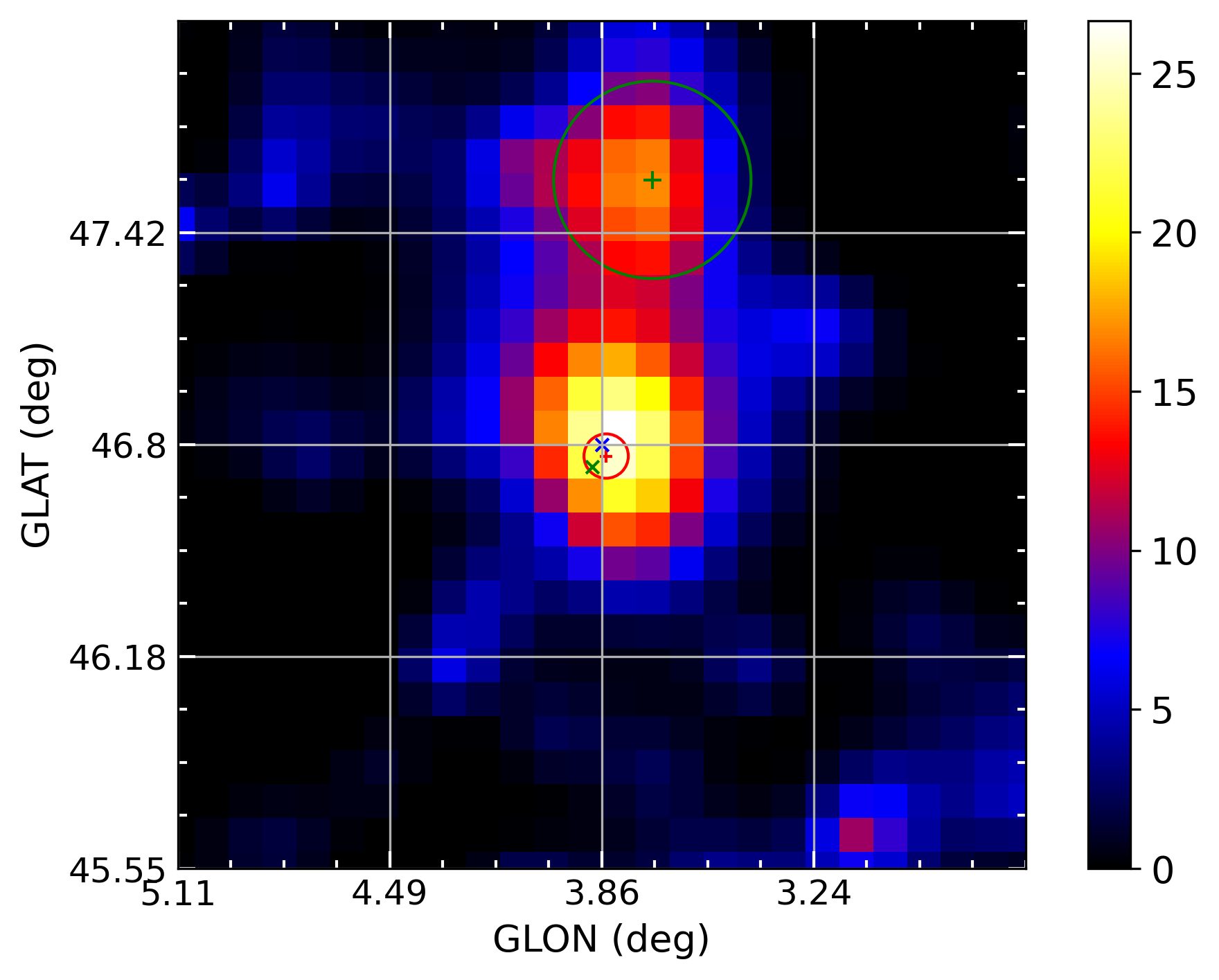}
        \caption{Left: $1^\circ\times 1^\circ$ TS excess map of M5 (in Galactic coordinates) in the 1$-$500 GeV band with a bin size of $0\fdg05$. Overlaid are the LAT best localization and 95\% localization uncertainty in this work (red plus sign and circle), 4FGL-DR4 position and 95\% localization uncertainty (green cross and ellipse), and M5 center (blue cross). Right: $2\fdg5 \times 2\fdg5$ TS excess map of M5 in the 0.1$-$500 GeV band with a bin size of $0\fdg1$. In addition to the markers for M5, the green plus sign and circle show the best localization and 95\% localization uncertainty for the excess in bin 21 of the 90-day light curve (see Section \ref{lightcurve} and Figure~\ref{fig:lc}).}
    \label{fig:tsmap}
\end{figure*}

\subsection{Localization}
Since the LAT cannot resolve M5, we modeled it as a point source in our analysis. We first localized it in the higher energy range of $1-500$ GeV to take advantage of LAT's better spatial resolution.
Data were binned using a $0\fdg05 \times 0\fdg05$ pixel size and 10 logarithmic energy bins per decade. 
The ROI was optimized by fitting all sources in the ROI to ensure that all parameters are close to their global likelihood maxima. 
We tested three spectral models for M5 during the localization: the default LogParabola (LP) model in 4FGL-DR4,
\begin{equation}
\frac{dN}{dE}=N_0\left(\frac{E}{E_0} \right)^{-(\alpha+\beta \ln(E/E_0))} \, \, \, ,
\end{equation}
the simple power-law (PL) model,
\begin{equation}
\frac{dN}{dE}=N_0\left(\frac{E}{E_0} \right)^{-\Gamma_{0}} \, \, \, ,
\end{equation}
and the PLSuperExpCutoff4 (PLEC4) model that is used for pulsar emission in the Third Fermi-LAT Catalog of Gamma-ray Pulsars \citep[3PC;][]{3PC},
\begin{equation}
\frac{dN}{dE}=N_0\left(\frac{E}{E_0} \right)^{-\Gamma+d/b} exp\left[ \frac{d}{b^2} \left( 1-\left(\frac{E}{E_0}\right)^b \right) \right] \, \, \, ,
\end{equation}
where $N_0$ is the normalization; $\alpha$ and $\Gamma_0$ are the spectral indices in the LP and PL models, respectively; $\beta$ is the curvature index; $E_0$ is the reference energy and is fixed to the catalog value for the LP model and to 1 GeV for the PL and PLEC4 models; $\Gamma$ is the local spectral index at $E_0$ in the PLEC4 model; $d$ is the local curvature at $E_0$; and $b$ is fixed to $2/3$.

Only the normalizations of the Galactic/isotropic diffuse components were set free to vary during the localization. 
%
%
%
We note that changing the spectral model and leaving the position free simultaneously in each fit will lead to ``un-nested models''.
However, the localizations determined from the three models are completely consistent (Table~\ref{tab:latfit}).
%
Both PLEC4 and LP models have the same number of degrees of freedom and fit the data quite well. Since the TS values obtained with both models are basically the same, we chose to continue our analysis with the PLEC4 model simply because it is physically motivated as a superposition of curvature radiation spectra for a range of electron energies 
\citep{Bednarek2007,Venter2008,Venter2009,Cheng2010}.
%

Figure~\ref{fig:residualmap} (left panel) shows the residual map for the full ROI for the localization generated by removing the contribution of all 4FGL sources in the ROI. No significant residuals ($>4\sigma$) were found, indicating a good modeling of the ROI.
The TS excess map (Figure~\ref{fig:tsmap}, left panel) presents the localization result in a $1^\circ\times 1^\circ$ region centered at M5 by removing the contribution of all 4FGL sources except M5, i.e., all sources except M5 are included in the model.

\subsection{Spectral fit}

\begin{table*}[htbp]
\scriptsize
\begin{center}
\caption{Fermi-LAT Analysis Results for M5}
\label{tab:latfit}
\begin{tabular}{lccccccc}
\toprule %
\multicolumn{8}{c}{$1-500$ GeV Localization} \\
\midrule %
Model   & TS  &GLON    &GLAT      & 95\% Localization Uncertainty   \\
        &    & (deg)   & (deg)    &  (deg)  \\
\midrule %
LP        & 73.3   & $3.8458\pm0.0253$   & $46.7606\pm0.0282$  & 0.07  \\
PL      & 71.1   & $3.8409\pm0.0261$   & $46.7598\pm0.0286$  & 0.07  \\
PLEC4    & 73.8   & $3.8453\pm0.0254$   & $46.7622\pm0.0283$  & 0.07  \\

\midrule %
\multicolumn{8}{c}{$0.1-500$ GeV Spectral Fits} \\
\midrule %
  &	 TS	 & $\alpha$  & $\beta$  & $\Gamma$ ($\Gamma_{0}$) & $d$ & Photon Flux  & Energy Flux\\
   &   &      &   &   &   &(10$^{-9}$~cm$^{-2}$~s$^{-1}$)   & (10$^{-12}$~erg cm$^{-2}$~s$^{-1}$)  \\
\midrule %
LP  &90.9	   &$2.11\pm0.20$ & $0.44\pm0.18$	& ... & ...	&$1.41\pm0.60$ & $1.75\pm0.32$	\\
PL  &74.7	& ...  & ...  &$2.24\pm0.09$	& ... 	&$4.02\pm0.89$ & 
$2.90\pm0.41$	\\
PLEC4  &91.2	& ...   & ...  &$1.70\pm0.25$	& $0.63\pm0.24$	&$1.65\pm0.67$ & 
$1.78\pm0.32$	\\
\bottomrule %
\multicolumn{8}{p{17cm}}{{\bf Notes}: LP stands for LogParabola, PL for power law and PLEC4 for PLSuperExpCutoff4. $\Gamma_{0}$ is the spectral index for the PL model, and $\Gamma$ is the local spectral index for the PLEC4 model.} \\
\end{tabular}
\end{center}
\end{table*}

We performed a broadband spectral fit for M5 in the energy range of 0.1$-$500 GeV using the best-fit localization obtained with the PLEC4 model.
Data were binned using a $0\fdg1 \times 0\fdg1$ pixel size and 10 logarithmic energy bins per decade.
This time, all the spectral parameters of sources within $5^\circ$ and those of the Galactic/isotropic diffuse components were set free to vary, along with the normalizations of significantly variable sources within $5^\circ-8^\circ$ around M5.

Similarly, we tested the three aforementioned spectral models, and we summarize the best-fit results in Table~\ref{tab:latfit}.
The PLEC4 and LP models are indistinguishable, while the PL model is the worst. We again took PLEC4 as the best-fit model based on its physical motivation.
Figure~\ref{fig:residualmap} (right panel) shows the residual map for the full ROI for the broadband fit in which no significant residuals ($>4\sigma$) are present.

\begin{figure}[htbp]
    \centering 
        \includegraphics[width=1.0\linewidth]{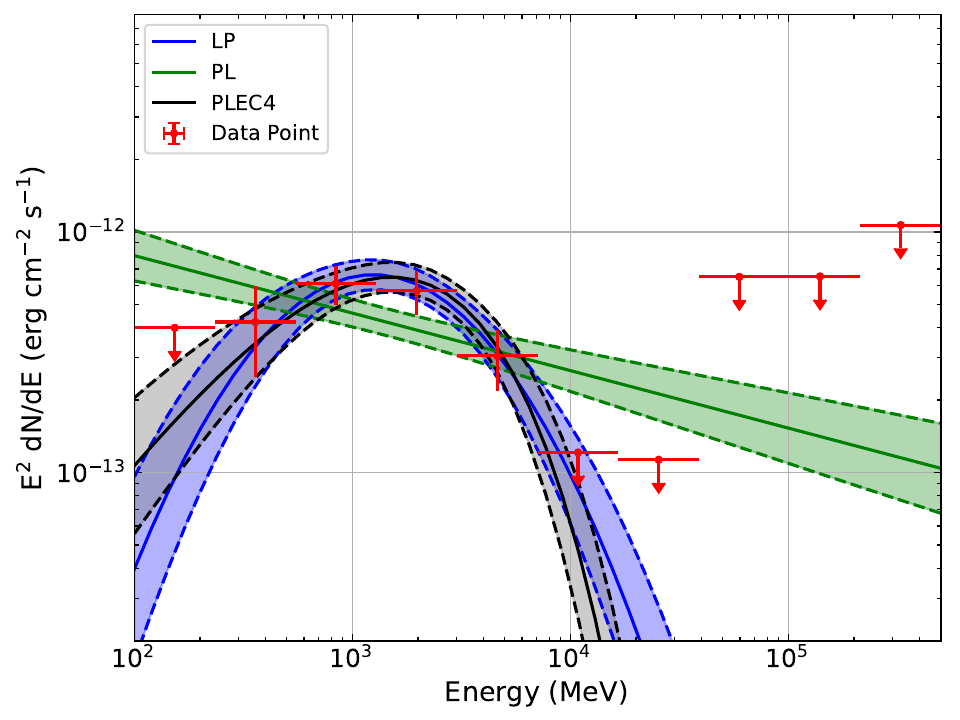}
        \caption{Gamma-ray spectrum of M5 (red circles), along with the broadband models of PLEC4 with uncertainty (black line and shaded region), LP with uncertainty (blue line and shaded region), and PL with uncertainty (green line and shaded region). Flux upper limits at the 95\% CL are shown as red arrows for bins when the source had TS $<$ 4.}
    \label{fig:sed}
\end{figure}

Taking the best-fit PLEC4 model in $0.1-500$ GeV, we computed the spectral energy distribution (SED) by performing a maximum likelihood analysis in 10 logarithmically spaced energy bins over 0.1$-$500 GeV. Background sources that were free in the broadband fit were kept free in the fit of each bin.
The normalization of M5 in each bin is fit using a PL spectral parameterization with a fixed index. At lower energy bands (the first six bins), the index was fixed to the local slope of the broadband PLEC4 model. At higher energy bands (the last four bins), the index was fixed to 4 in order to be consistent with the local slope used at lower energies, given that the PLEC4 model is both very steep and very low in normalization at higher energies. 
Upper limits on the flux at the 95\% confidence level (CL) were computed for bins when the source had TS $<$ 4. 
Figure~\ref{fig:sed} shows the SED, along with the PLEC4, LP, and PL models.

\subsection{Gamma-ray Pulsation Search}
\label{pulsation}

A gamma-ray pulsation search for each pulsar has been performed by selecting photons within $2^\circ$ of the pulsar position. 
Spin phases of gamma-ray photons were calculated using the Fermi plug-in \citep{Ray2011} for \texttt{TEMPO2} and the radio ephemerides for each pulsar \citep{lZhang2023}, respectively. The ephemerides for M5A, N5B, M5C, M5D, and M5E are valid from before Fermi was launched to 2022 November, while those for M5F and M5G are valid for the time range of 2020 November 16$-$2022 December 14, since these two MSPs were the newest in M5 and only detected by FAST.
We used the weighted H-test \citep{kerr2011} statistic to quantify the pulsation significance, with weights computed by employing the Simple Weights method as described in \cite{Bruel2019} and \cite{Smith2019} for both the full LAT data set and that in the time range of the ephemerides validity. 
No significant pulsations have been detected from any of the seven MSPs. The largest H-test value is 13.8 found for M5A corresponding to around 2.9$\sigma$, which decreases slightly when considering the six trials used in the search.

\subsection{Gamma-Ray Variability}

\subsubsection{Long-term Light Curves}
\label{lightcurve}
\begin{figure*}[htbp]
    \centering 
        \includegraphics[width=0.48\linewidth]{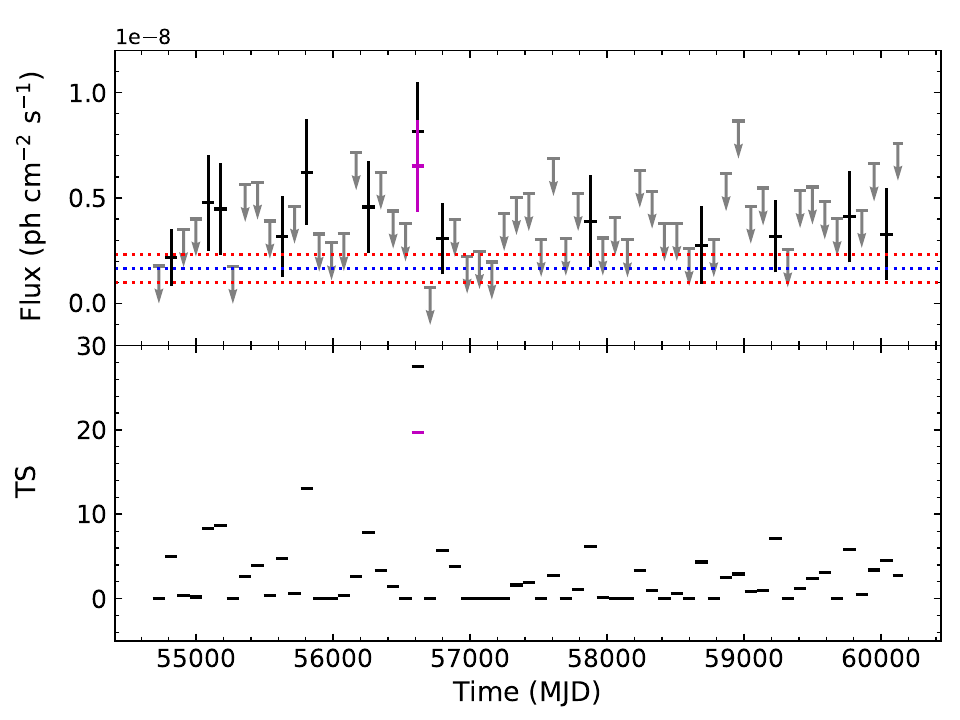}
        \includegraphics[width=0.48\linewidth]{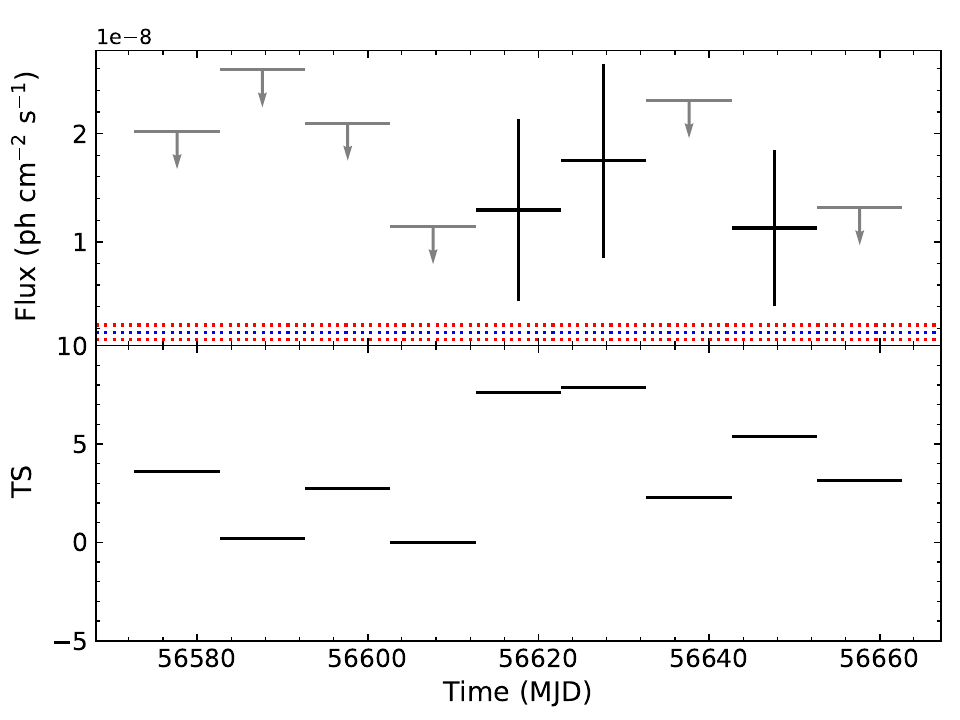}
        \caption{Long-term light curve and TS evolution for M5 in the energy range of 0.1$-$500 GeV. Left: 90-day binning. Right: a zoom-in around bin 21 of the 90-day light curve with a binning of 10 days (from MJD 56572 to MJD 56662). Flux upper limits at the 95\% CL are shown as arrows for bins when the source had TS $<$ 4. Blue dotted line: average flux in the broad band. Red dotted lines: 1$\sigma$ uncertainties on the average flux. The magenta marker indicates the flux and TS in bin 21 after taking into account the excess near M5.}
    \label{fig:lc}
\end{figure*}

To investigate the long-term gamma-ray flux variability of M5, we computed a light curve with a 90-day binning over $0.1-500$ GeV (Figure~\ref{fig:lc}, left panel). 
The best fit of the PLEC4 model in the 0.1$-$500 GeV band obtained previously was used as the starting point for each time bin, but this time only the normalizations were kept free. 
An independent binned likelihood analysis was performed for each bin to get the flux of M5.
Upper limits at the 95\% CL were calculated when M5 had TS $<$ 4. 
We followed the same method as presented in \cite{3FGL} to quantify the variability significance and obtained TS$_{\rm var}=62.97$. In a $\chi^2$ distribution with 60 degrees of freedom, the 99\% confidence TS threshold above which the variability would be considered probable is 88.38. 
Thus, the gamma-ray emission from M5 is consistent with being steady on a time scale of a few months.

However, there is a TS peak of $\sim28$ in the time bin 21 spanning MJD 56572$-$56662 (2013 October 7$–$2014 January 5). To further investigate the peak, we computed a TS map for this bin with M5 removed from the source model (Figure~\ref{fig:tsmap}, right panel).
A potential excess appeared near M5. We then localized this excess to $(l, b)=(3\fdg71, 47\fdg58)$ with a 95\% localization uncertainty of $0\fdg29$, which, despite being large, does not enclose M5 ($0\fdg78$ offset from the M5 center).
Therefore, this excess is not related to M5. We verified that the excess does not appear in the bins before and after bin 21. 
A likelihood fit in this bin with the excess added 
resulted in a TS of around 20 and 10 for M5 and the excess, respectively. 
It is interesting to note that a TeV source that is offset by $4^\prime$ from the center of the GC Terzan 5 and extended well beyond the HESS PSF was detected with unconfirmed origin \citep{HESS2011}. 

%
We have searched multiple catalogs for possible counterparts within the 95\% localization uncertainty of the excess. We first checked the Fermi LAT Long-Term Transient Source Catalog \citep[1FLT;][]{Baldini2021}\footnote{\url{https://heasarc.gsfc.nasa.gov/W3Browse/all/fermiltrns.html}} which contains transient sources detected above $4\sigma$ on monthly time scales using 10 yr of Fermi-LAT data.
Given the low significance of the excess, we expected to find no counterpart in the 1FLT catalog, and indeed there is none.
Then, through HEASARC,\footnote{\url{https://heasarc.gsfc.nasa.gov/W3Browse/all/}} we searched for blazar candidates in the FERMILBLAZ \citep{fermiblz2019}, BZCat \citep{Massaro2015}, CGRaBS \citep{Healey2008}, CRATES \citep{Healey2007}, WIBRaLs2, and KDEBLLACs catalogs \citep{Abrusco2019,Menezes2019}}, but no matches were found for the excess.
We note as well that, given the high Galactic latitude of M5, it is highly unlikely to find novae, and indeed no novae were reported in the relevant time period.\footnote{\url{https://asd.gsfc.nasa.gov/Koji.Mukai/novae/novae.html}}

After adding the potential excess to the source model, the TS$_{\rm var}$ of the 90-day light curve (Figure~\ref{fig:lc}, left panel) decreased to 57.9.
Nevertheless, the TS of M5 in bin 21 after taking the excess into account remains an outlier, although TS is not a good measure of variability (it can vary owing to nearby sources or just exposure).
To quantify whether the apparent high TS of M5 in bin 21 is significant or not, we compared the best-fit $\log \mathcal{L}$ in this bin with the $\log \mathcal{L}$ of the fit when fixing the flux of M5 to the average, both with the excess added in this bin.
We then computed $\sqrt{2\Delta \log \mathcal{L}}$, i.e., Sqrt\_TS\_History in the 4FGL catalog, which gives the significance in $\sigma$ units for 1 degree of freedom. We obtained 2.8$\sigma$, moderately significant. The apparent TS peak in bin 21 is thus compatible with being a statistical fluctuation. 
Besides, we zoomed in on  bin 21 by computing a 10-day light curve (Figure~\ref{fig:lc}, right panel) for it. The TS$_{\rm var}$ is 15.2, while the 99\% confidence TS threshold with 8 degrees of freedom is 20.1. 
Thus, the gamma-ray emission inside bin 21 is compatible with a constant signal and is also compatible with bin 21 being a positive statistical fluctuation as concluded above. 
%

\subsubsection{Orbital Modulation}

\begin{figure*}[htbp]
    \centering 
        \includegraphics[width=0.48\linewidth]{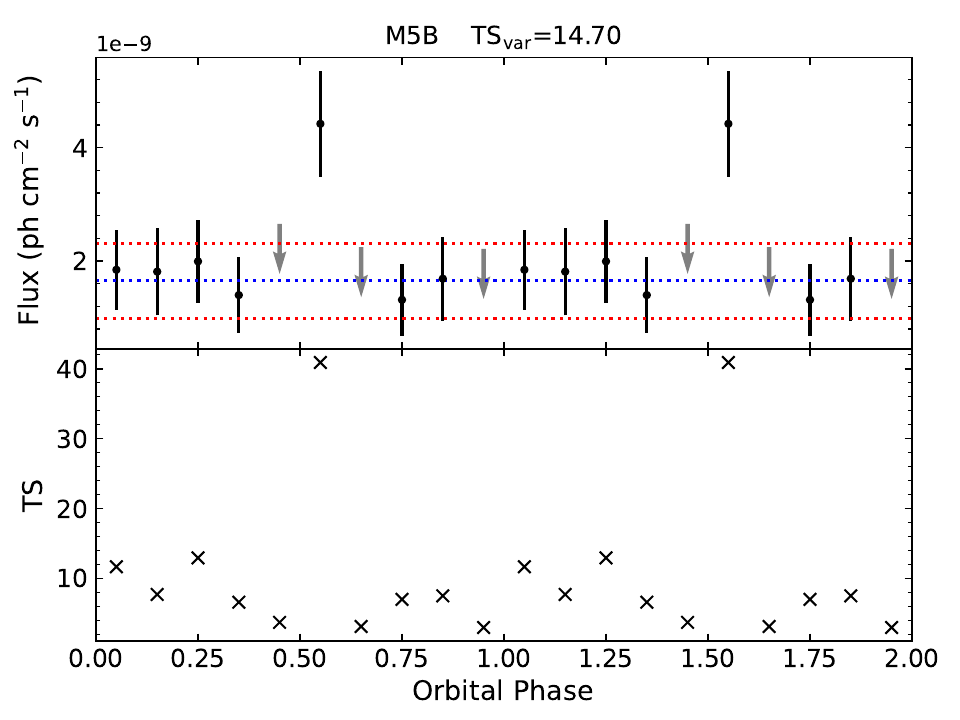}
        \includegraphics[width=0.48\linewidth]{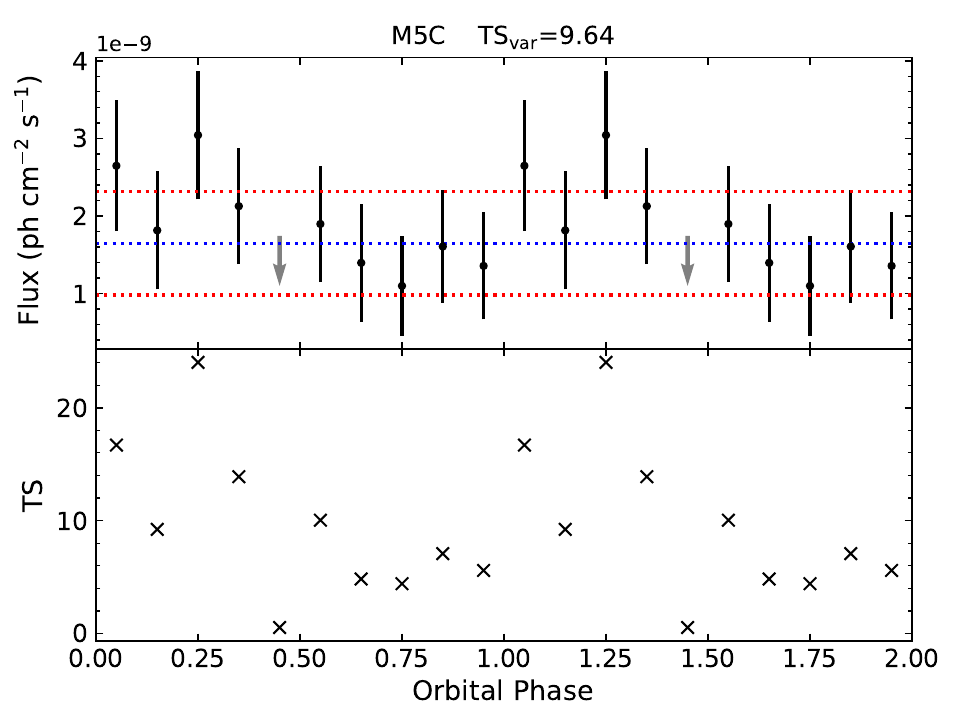}
        \includegraphics[width=0.48\linewidth]{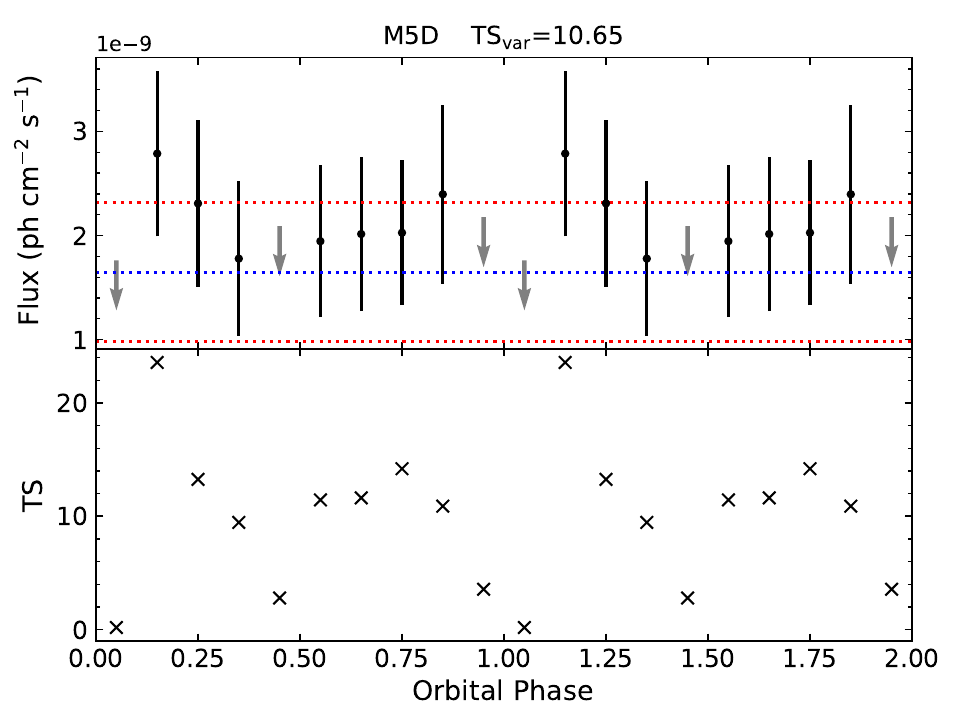}
        \includegraphics[width=0.48\linewidth]{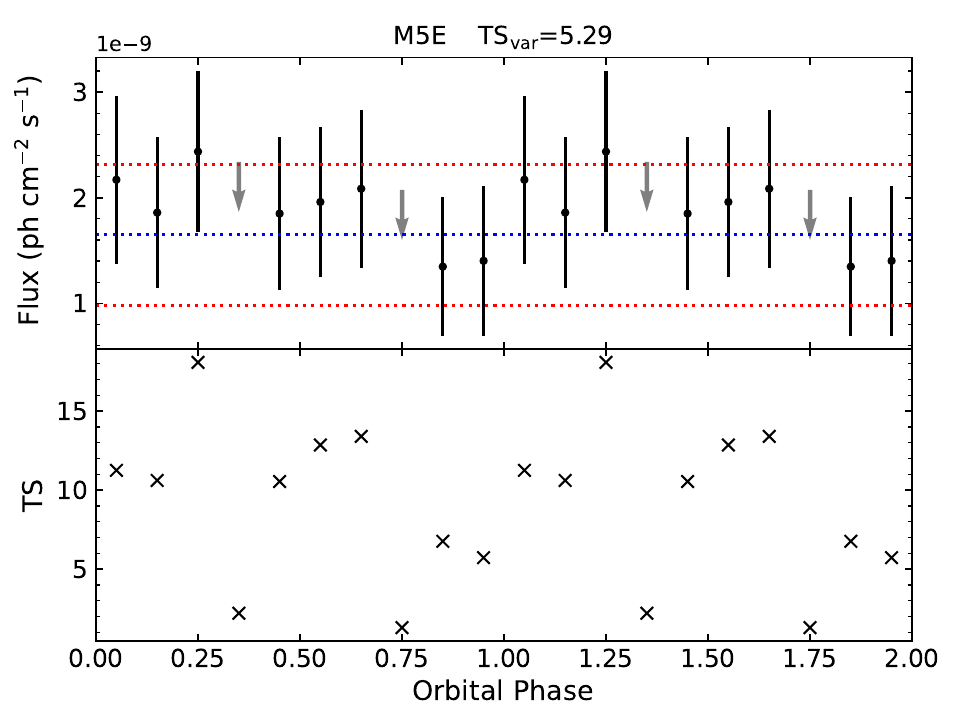}
        \includegraphics[width=0.48\linewidth]{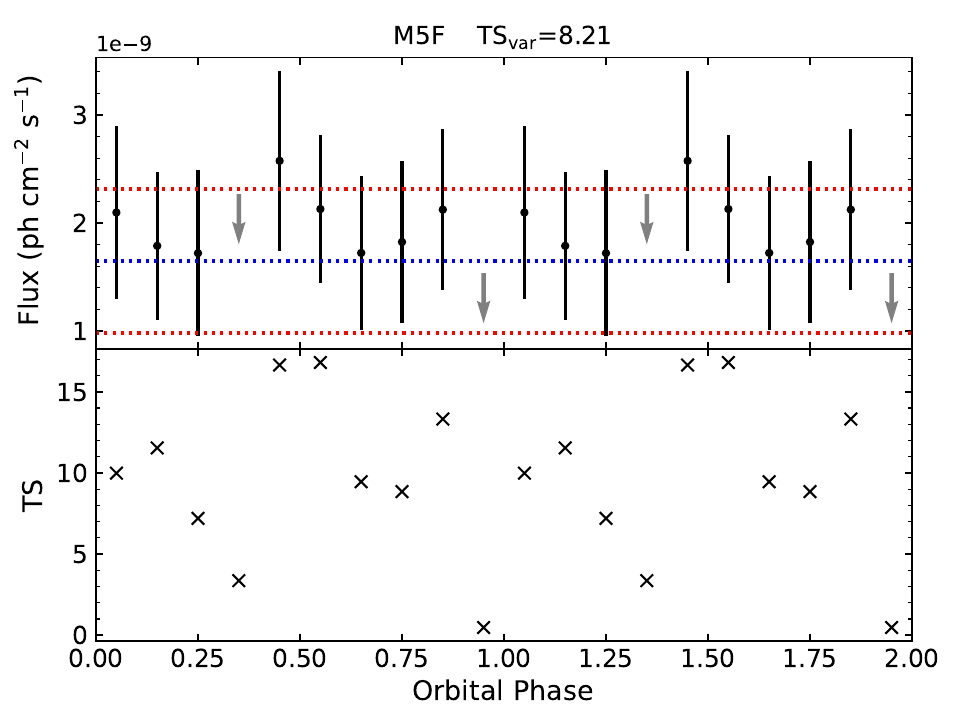}
        \includegraphics[width=0.48\linewidth]{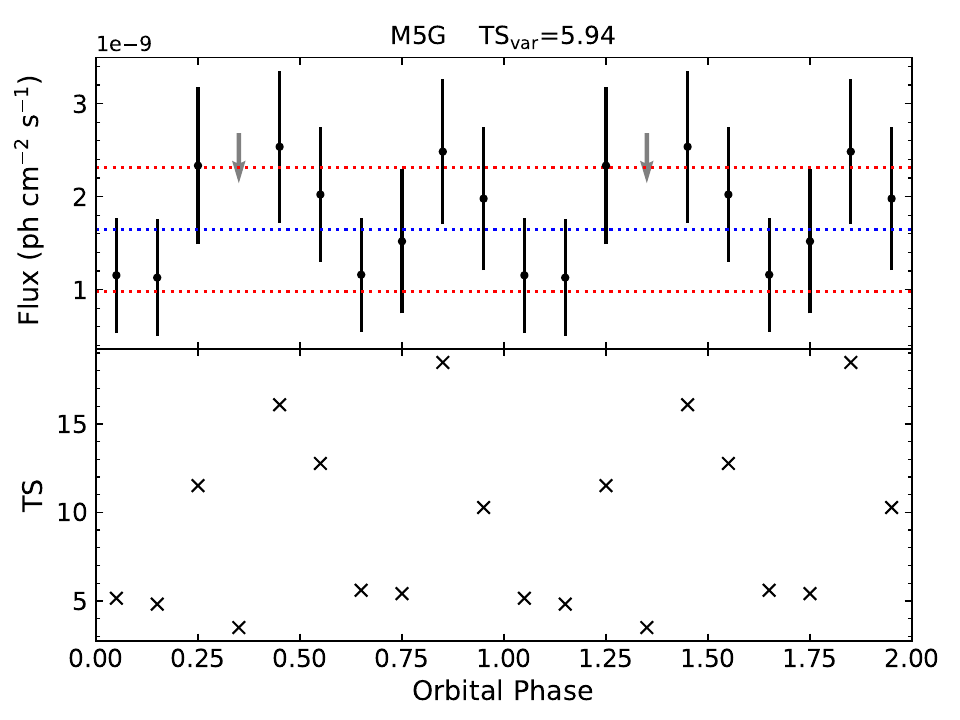}
        \caption{Orbital flux modulation for the six binary MSPs in M5. Upper limits at the 95\% CL are shown as arrows for bins when the source had TS $<$ 4. Blue dotted line: average flux in the broadband fit with PLEC4 model. Red dotted lines: 1$\sigma$ uncertainties on the average flux.}
    \label{fig:orbi_lc}
\end{figure*}

Binary MSPs can exhibit orbitally modulated emission in multiple wavelengths such as optical, X-ray, and gamma rays. Detecting modulation constrains the binary system properties and theoretical models for the multiwavelength emission of such systems.
Orbital phases were assigned to each event using the Fermi plug-in \citep{Ray2011} for \texttt{TEMPO2} and the radio timing solution for each pulsar \citep{lZhang2023}, respectively. 
We adopted two approaches to study the orbital modulation of the six binary MSPs in M5, similar to what was done in \cite{Johnson2015}. 
We first created a counts light curve of $5^\circ$ around the M5 center with a 30 s time bin and folded it with the orbital period of each pulsar. 

In the first approach, we calculated the LAT exposure for each time bin in the counts light curve to account for any possible orbital variations of the exposure \citep{Ackermann2012}. 
By binning the exposure in 1000 bins of orbital phase and normalizing, we built a null distribution of what the orbital modulation should look like if the exposure variation versus orbital phases is the only ``modulation'' present. 
We then used this null distribution to get the exposure-corrected orbital phase for each event in a region of $2^\circ$ around M5. 
Finally, we employed the weighted H-test \citep{kerr2011} to quantify the orbital modulation for each pulsar. 
Weights were calculated in a similar way to that in
the gamma-ray pulsation search (Section \ref{pulsation}). 
No strong evidence of modulation was found for any of the six MSPs. The largest H-test value found was 10.3 for M5B corresponding to 2.4$\sigma$, without trial corrections.

In the second approach, we computed the orbital flux for each MSP with 10 bins per orbit following the same methodology as for the long-term light curve presented above. 
One extra step before performing the binned likelihood fit in each orbital bin is to correct the potential LAT exposure variations across the orbit by creating orbital-phase-selected good time intervals from the orbital-period-folded 30 s count light curve that was generated previously. 
Similar to the long-term light-curve analysis, we computed TS$_{\rm var}$ to quantify the flux variability. 
The orbital flux modulations using this approach are shown in Figure~\ref{fig:orbi_lc}. 
The $\chi^2$ distribution with 9 degrees of freedom corresponds to a 99\% confidence TS threshold of 21.7. None of the six MSPs has TS$_{\rm var}$ larger than this value, with the largest being 14.7 found for M5B. Thus, no significant variability was found for them.
Despite M5B's apparently high TS and flux in the orbital bin of 0.5-0.6 (Table~\ref{tab:orbifit_M5B} and Figure~\ref{fig:orbi_lc}), the variability over the full orbit is not significant.

Similar to the long-term light-curve analysis, we computed $\sqrt{2\Delta \log \mathcal{L}}$ to quantify whether the high TS and flux in the bin of 0.5-0.6 of M5B are significant compared to the phase-averaged flux (Table~\ref{tab:latfit}). We obtained 3.4$\sigma$. Taking into account the 60 trials (number of orbital bins 10 multiplied by the number of pulsars 6), the probability to get a 3.4$\sigma$ excess is 4\%, corresponding to 2$\sigma$ after trial corrections. The flux and TS variation of M5B are thus not significant and are probably due to statistical fluctuations.
%
To further investigate the orbital modulation, we then compared the spectral characteristics in the orbital bin of 0.5-0.6 and 0.6-0.5 by reperforming likelihood fits in the two bins.
This is done similarly to the standard orbital modulation analysis described above, but this time the spectral shape parameters of M5 were also set free to vary in order to test any shape changes in addition to the overall flux variation.
The results are presented in Table~\ref{tab:orbifit_M5B}. 
The flux difference between 0.5-0.6 and 0.6-0.5, as well as the phase-averaged fit (Table~\ref{tab:latfit}), is less than $\sim 3\sigma$, consistent with being a statistical fluctuation as stated above. The spectral shape parameters are, on the other hand, consistent within uncertainties.

\begin{table*}
\scriptsize
\begin{center}
\caption{Orbital-phase resolved Spectral Fits for M5B}
\begin{tabular}{lccccc}
\toprule %
 Bin  & TS       & $\Gamma$     & $d$         &Photon Flux      & Energy Flux\\
     &    &  &  &  (10$^{-9}$~cm$^{-2}$~s$^{-1}$)   & (10$^{-12}$~erg cm$^{-2}$~s$^{-1}$)  \\
\midrule %
0.5-0.6 &42  & $2.08\pm0.33$	& $0.61\pm0.38$   & $7.70\pm4.67$  & $5.70\pm1.70$	\\
0.6-0.5 &61	 & $1.72\pm0.33$	& $0.56\pm0.28$   &$1.50\pm0.84$   & $1.58\pm0.39$	\\
\bottomrule %
\label{tab:orbifit_M5B}
\end{tabular}
\end{center}
\end{table*}

\section{Discussion and Concluding Remarks}
\label{discuss}

From the spectral parameter values we obtained in the fit, we can evaluate the energy at which the SED peaks as 
\begin{equation}
E_{p}=E_{0}\left[(1+\frac{b}{d}(2-\Gamma)\right]^{\frac{1}{b}} \, \, ,
\end{equation}
and the curvature at the SED peak $d_{p}$ as 
\begin{equation}
d_{p}=d+b(2-\Gamma)  \, \, ,
\end{equation}
as outlined in 3PC \citep{3PC}, with $d_{p}$ reaching a maximum of $4/3$ for synchrotron or curvature radiation from monoenergetic electrons. $E_{p}$ and $d_{p}$ are correlated.
The SED peak width is inversely proportional to $d_{p}$ such that high curvature indicates a narrow spectrum corresponding to a narrow range of electron energies and low curvature indicates a broad spectrum with contributions from a broader range of electron energies.
We obtained $E_{p}=1.5$ GeV and $d_{p}=0.83$ for M5, putting it near the upper right corner of the $d_{p}$ vs. $E_{p}$ plot (see Figure 20 in the 3PC). Following the 3PC and taking into account the values we obtained, if the emission detected from the GC comes mostly from a single source, or if it comes from many sources behaving similarly, the electron population producing the emission is rather broad.

\subsection{Nondetection of Individual Pulsars and Implications}

While 305 radio pulsars in GCs are currently known, with most of them being MSPs, only three individual GC MSPs have been detected in gamma rays, or about 1\%. This is partly due to the large distances of GCs, which imply that any MSPs detected must be exceptionally energetic (especially in the earlier phases of the Fermi mission: the first two GC MSPs detected in gamma rays are by far the most powerful known, with spin-down power $\dot E> 8 \times 10^{35}$~erg~s$^{-1}$). However, this might also be due to the fact that not many GC pulsars have ephemerides that are accurate for a large fraction of the Fermi mission's 15 yr. Thus, whenever such ephemerides become available, it is important to verify whether the pulsars can be detected in gamma rays: any such detection would automatically imply an exceptionally energetic MSP.
%

From Table~\ref{tab:M5info}, we can see that some of the MSPs in M5 have $\dot E$ upper limits that are comparable with not only those of gamma-ray MSPs \citep{3PC} but even those of gamma-ray MSPs detected in GCs: for instance, M5C has an upper $\dot{E}$ limit of $1.2 \times 10^{35} \rm \, erg \ s^{-1}$, while for J1835$-$3259B (where we can estimate the intrinsic spin-down from
the measured spin-down and the variation of the orbital period reported by \citealt{2022A&A...664A..54G}) $\dot{E} = 1.8 \pm 1.2 \times 10^{35} \rm \, erg \ s^{-1}$. Furthermore, the detectability of the pulsars in M5 is enhanced by the fact that the photon background toward M5 is significantly smaller than that toward NGC 6652, whose gamma-ray emission is probably dominated by a single pulsar.

The nondetection suggests that the true values of $\dot{E}$ for the M5 pulsars are well below the upper limits derived by \cite{lZhang2023}.
This is not surprising given previously observed trends among the GC pulsars. The three pulsars detected in gamma rays are all located in GCs with a very high $\gamma_{b}$. Two of them (J1823$-$3021A and J1835$-$3259B) are located in core-collapsed GCs, NGC~6624 and NGC~6652, which generally have the highest values of $\gamma_{b}$ \citep{Verbunt_Freire2014}.

The latter authors have remarked that most of the relatively slow (thus higher magnetic field $B$) pulsars are located in high-$\gamma_{b}$ clusters. They ascribed this to the high stellar encounter rate itself: this will disrupt many pulsar binaries, leading to a high rate of isolated pulsars in core-collapsed GCs, which has been repeatedly confirmed \citep[e.g.,][]{Abbate+2022,Abbate2023}.
This high encounter rate could also be responsible for the prevalence of slow pulsars in these clusters if LMXBs are also being disrupted, leaving behind partially recycled NSs, which will appear as slower radio pulsars with larger $B$-fields than Galactic MSPs. It is possible that B1821$-$24A and J1823$-$3021A were formed in this way: although they were spun up, their $B$-fields had not been fully ablated, resulting in the very large magnetic braking torque and the unusually large $\dot{E}$. Nevertheless, the disruption of compact binary systems in GCs by close stellar encounters is still an open topic of research and was recently contested by \cite{Menezes2023}. These authors used the Heggie-Hills law \citep{Heggie1975,Hills1975} for binary encounters in combination with Fermi-LAT and Chandra data to argue that compact binary ionization would happen only in the unrealistic scenario where the dispersion velocity of stars in the cores of GCs is greater than the GCs' escape velocity.

In GCs like M5, with much smaller values of $\gamma_{b}$, any LMXBs, once formed (in exchange interactions), are not likely to be disturbed again, recycling their respective NSs right through to the end. Thus, we would expect all pulsars in these clusters to not have only fast spins but also small $B$-fields, as generally observed for MSPs in the Galactic disk.  This is confirmed in 47 Tuc, where we can say that all MSPs in binaries (for which the cluster acceleration can be estimated from precise measurements of variation of the orbital period) have small values of $\dot{P}$, similar to those of Galactic MSPs \citep{Freire+2017}. This is confirmed by the fact that none of the 23 pulsars with long-term timing solutions in 47~Tuc, or the fewer known in $\omega$ Centauri, are individually detectable in gamma rays \citep{Dai2023}. For the same reasons, we do not really expect the occurrence of pulsars with high $B$-fields in low-$\gamma_{b}$ GCs like M5.

Nevertheless, attempting to detect powerful gamma-ray MSPs in low-$\gamma_{b}$ GCs like M5 serves as a useful test of these ideas. If we find a single bright gamma-ray MSP in these clusters, we will know that their formation is not necessarily linked to processes that occur almost exclusively in high-$\gamma_{b}$ GCs, like LMXB disruption.

\subsection{Gamma-Ray Emission of the Cluster as a Whole}

Following the above discussion, the gamma-ray emission from M5 is therefore the resulting collective emission of individual pulsars in the cluster, similar to the findings in the study of 47~Tuc and $\omega$ Centauri. 
We can thus follow the method presented in \cite{Johnson2013} to estimate the total number of MSPs in M5 as
\begin{equation}
    N_{\rm {MSP}} = \frac{L_{\gamma}}{\langle \dot E\rangle \langle \eta_{\gamma}\rangle} \,\,\, ,
\end{equation}
where $\langle \dot E\rangle =(1.8\pm0.7)\times10^{34}$~erg~s$^{-1}$ is the average spin-down power in GCs \citep{Abdo2009}, 
$\langle \eta_{\gamma}\rangle$ is the average gamma-ray efficiency of MSPs, and $L{\gamma}=(1.2\pm0.2\pm0.2)\times10^{34}$ ~erg~s$^{-1}$ is the gamma-ray luminosity\footnote{The first uncertainty on $L{\gamma}$ comes from the statistical uncertainty in the spectral fit, and the second one is systematic induced by the distance uncertainty.} of M5 based on the spectral fit reported in Table~\ref{tab:latfit}.
%
%
%
For $\langle \eta_{\gamma}\rangle$, 3PC includes 20 MSPs with proper-motion and distance measurements yielding systematic uncertainties on $\eta_{\gamma}$, after Shlovskii corrections, smaller than 50\%, similar to the criteria in \cite{Johnson2013}.
Instead of $\langle \eta_{\gamma}\rangle$, we used the FWHM range $0.07< \eta_{\gamma} < 0.4$ of the $\eta_{\gamma}$ distribution, as in Figure 24 of the 3PC, but with Shlovskii corrections for only the 20 well-characterized MSPs. This gives $1.7 < N_{\rm {MSP}}< 9.5$.
The current number of seven known MSPs in M5 is in this range. There may be a few more (but not a large number of) MSPs to discover in M5.

%

%
However, such estimates are to be taken with care, as we may see (or not) a pulsar just based on geometry. Measured efficiency can be severely affected by orientation. 
3PC assumes a beaming factor $f_{\Omega} = 1$, while recent theoretical studies of LAT pulsars found that in general $f_{\Omega}<1$ is expected \citep{Kalapotharakos2022}, adding another uncertainty on the estimates.

Although it is generally believed that the gamma-ray emissions of GCs are from MSPs, there are two main distinct models.
The pulsar magnetosphere model proposes that gamma rays are produced via curvature radiation of relativistic electrons/positrons in the pulsar magnetosphere. The gamma rays from GCs are thus expected to originate from the cumulative contribution of all MSPs in the cluster \citep[e.g.,][]{Venter2008,Venter2009}. 
%
The inverse Compton (IC) model, on the other hand, suggests that gamma rays are generated by the IC scattering between relativistic electrons/positrons in the pulsar wind of MSPs in GCs and background soft photons \citep{Bednarek2007,Venter2009,Cheng2010}, which will lead to intrinsic unpulsed emission of GCs.

Currently, both models can explain the GeV gamma-ray spectra of GCs equally well \cite[e.g.,][]{Abdo2010,Cheng2010}. The IC model, on the other hand, also predicts TeV gamma rays. However, observations with CANGAROO III, VERITAS, H.E.S.S, and MAGIC of GCs have not been successful \cite[see e.g.,][]{Kabuki2007,Aharonian2009,Anderhub2009,McCutcheon2009,HESS2013,MAGIC2019}, with Terzan 5 being the only one to be claimed to shine in the TeV band \citep{HESS2011}.
Diffuse radio and X-ray emissions from GCs can also be produced by synchrotron radiation and IC scattering, as predicted by the IC model \citep{Cheng2010}. Observational support for such a scenario has been provided by the discovery of extended radio and X-ray emissions around Terzan 5 \citep{Eger2010,Clapson2011} and 47 Tuc \citep{Wu2014} with possibly nonthermal origin.
See \cite{Tam2016} for a review of the observations and modelings of gamma-ray emission from GCs.

\cite{Song2021} claimed to have found evidence of a power-law high-energy tail in the gamma-ray spectra of GCs beyond the exponential cutoff power-law component by analyzing Fermi-LAT data for 157 MW GCs, which they interpreted in terms of the IC model, although more data are needed to assure their findings.
They also claimed that the very soft high-energy tail is the reason behind the difficulty in detecting GCs with current TeV telescopes, and that it would be possible to detect GCs with more sensitive TeV telescopes such as CTA \citep{Actis2011} and LHAASO \citep{Cao2019}.


In conclusion, based on current observations, the most obvious possibility to explain the GeV gamma-ray emission of GCs is that it is produced by the collective emission of individual (but yet-undetected) gamma-ray pulsars in the cluster. 
However, further multiwavelength observations with more sensitive telescopes such as CTA \citep{Actis2011} and LHAASO \citep{Cao2019} in TeV, SKA \citep{Tan2015} in radio, and EP in X-rays \citep{Yuan2022}
are expected to provide better constraints on the two aforementioned emission models.

%

\acknowledgments
We thank the anonymous referee for the very useful comments and suggestions that helped to improve the paper. The Fermi-LAT Collaboration acknowledges generous ongoing support from a number of agencies and institutes that have supported both the development and the operation of the LAT, as well as scientific data analysis. These include the National Aeronautics and Space Administration and the Department of Energy in the United States; the Commissariat \`a l'Energie Atomique and the Centre National de la Recherche Scientifique/Institut National de Physique Nucl\'eaire et de Physique des Particules in France; the Agenzia Spaziale Italiana and the Istituto Nazionale di Fisica Nucleare in Italy; the Ministry of Education, Culture, Sports, Science and Technology (MEXT), High Energy Accelerator Research Organization (KEK), and Japan Aerospace Exploration Agency (JAXA) in Japan; and the K.~A.~Wallenberg Foundation, the Swedish Research Council, and the Swedish National Space Board in Sweden. Additional support for science analysis during the operations phase is gratefully acknowledged from the Istituto Nazionale di Astrofisica in Italy and the Centre National d'\'Etudes Spatiales in France. This work was performed in part under DOE contract DE-AC02-76SF00515.

The authors acknowledge support from the National Natural Science Foundation of China under grant Nos. 12373051, 12041303, 12273038, U2031117, and 11988101. X.H. is also supported by the National Key R\&D Program of China (2023YFE0101200). 
D.F.T is supported by grant PID2021-124581OB-I00 funded by MCIN/AEI/10.13039/501100011033 and 2021SGR00426 of the Generalitat de Catalunya, by the Spanish program Unidad de Excelencia María de Maeztu CEX2020-001058-M, and by MCIN with funding from European Union NextGeneration EU (PRTR-C17.I1).
W.Z. is supported by the National Scholarship Council (PhD fellowship from the China Scholarship Council (CSC) (No. 202107030003)).
%
%
P.W. is also supported by the Youth Innovation Promotion Association CAS (id. 2021055), the Cultivation Project for FAST Scientific Payoff and Research Achievement of CAMS-CAS, and the National SKA Program of China (No. 2020SKA0120200).
%
%
Work at NRL is supported by NASA.
This research has made use of the SIMBAD database, operated at CDS, Strasbourg, France.

\bibliography{M5}
\bibliographystyle{aasjournal}

\end{document}